\documentclass[12pt]{iopart}
\usepackage{iopams} 
\usepackage{epsf}

\begin{document}

\title[Weyl quantization and the Von Neumann uniqueness theorem]{Alternative linear structures associated with regular Lagrangians. Weyl quantization and the Von Neumann uniqueness theorem}

\author{E Ercolessi\dag\footnote{To whom correspondence should be addressed (ercolessi@bo.infn.it)}, A Ibort\ddag, G Marmo\S  ~and G Morandi$\|$}

\address{\dag Physics Department, University of Bologna, CNISM and INFN, Via Irnerio 46, I-40126, Bologna, Italy}

\address{\ddag Depto. de Matem\'aticas, Univ. Carlos III de Madrid, 28911 Legan\'es, Madrid, Spain}

\address{\S Dipartimento di Scienze Fisiche, University of Napoli and INFN, Via Cinzia, I-80126 Napoli, Italy}

\address{$\|$ Physics Department, University of Bologna, CNISM and INFN, V.le B. Pichat 6/2, I-40127, Bologna, Italy}

\begin{abstract}
We discuss how the existence of a regular Lagrangian description on the tangent bundle $TQ$ of some configuration space $Q$  allows for
the construction of a linear structure on $TQ$ that can be considered as ``adapted" to the given dynamical system. The fact then that many dynamical systems admit 
alternative Lagrangian descriptions opens the possibility to use the Weyl scheme to quantize the system in different non equivalent ways, "evading", so to speak,  the von Neumann uniqueness theorem.

\end{abstract}



\maketitle

\section{Introduction}

The main purpose of this Note is to discuss how the existence of a regular Lagrangian description $\mathcal{L}$ on the tangent bundle $TQ$ of some configuration space $Q$, can
allow, under suitable assumptions that will be discussed shortly below, for
the ``dynamical" construction of a linear structure on $TQ$ that can be
considered as ``adapted" to the given dynamical system. If and when this is
possible, one obtains a new action of the group $\mathbb{R}^{2n}$
$\left(  n=\dim Q\right)  $ on $TQ$ and, as will be shown, the Lagrangian
two-form $\omega_{\mathcal{L}}$ can be put explicitly in canonical Darboux form. One can then follow the Weyl procedure \cite{Weyl}  to quantize
the dynamics, by realizing the associated Weyl system on the Hilbert space of
square-integrable functions on a suitable Lagrangian submanifold of $TQ$.

The fact that many dynamical systems admit genuinely
alternative\footnote{I.e. not differing merely by the addition of a total time
derivative to the Lagrangian.} Lagrangian descriptions \cite{Ferr} poses an interesting question, namely: assume
that a given dynamical system admits alternative Lagrangian descriptions
with more than one regular Lagrangian. According to what has been outlined
above, one will possibly obtain different actions (realizations) of the group
$\mathbb{R}^{2n}$ on $TQ$ that in general will not be linearly related. Then,
it will be possible to quantize ``\`a la" Weyl the system in two different ways,
thereby obtaining different Hilbert space structures on spaces of
square-integrable functions on different Lagrangian submanifolds (actually
what appears as a Lagrangian submanifold in one scheme need not be such in the
other. Moreover, the Lebesgue measures will be different in the two cases).
The occurrence of this situation seems then to offer the possibility of,
so-to-speak,"evading" the von Neumann theorem \cite{Von} and this is one of the topics to be
discussed in this Note.

Before embarking in the general discussion, we recall here some known facts
about the possibility of defining alternative (i.e. not linearly related)
linear structures on a vector space and/or of using the linear structure of a
vector space to endow with a linear structure manifolds that are related to
the given vector space.

\section{Alternative linear structures}

Let  $E$ be a (real or complex) linear vector space. A (not
necessarily linear) diffeomorphism:%
\begin{equation}
\phi:E\leftrightarrow M
\end{equation}
with $M$ \ a manifold \ (possibly $M=E$) allows us to "import" a linear
structure from $E$ to $M$. In particular, if $M=E$, we can define an
alternative linear structure on $E$ itself. To do so, we proceed by defining:

\begin{itemize}
\item Addition of $u,v\in M$ as:%
\begin{equation}
u +_{\left(  \phi\right)  }v=:\phi(\phi^{-1}\left(  u\right)
+\phi^{-1}\left(  v\right)  ). \label{property1}%
\end{equation}

\item Multiplication by a scalar $\lambda\in\mathbb{R}$ or $\mathbb{C}$ of
$u\in M$ as:
\begin{equation}
\lambda\cdot_{\left(  \phi\right)  }u=:\phi\left(  \lambda\phi
^{-1}\left(  u\right)  \right) . \label{property2}%
\end{equation}

\end{itemize}

These operations have all the usual properties of addition and multiplication
by a scalar. In particular:
\begin{equation}
\left(  \lambda\lambda^{\prime}\right) %
\cdot_ {\left(  \phi\right)  } u=\lambda \cdot_ {\left(  \phi\right)  }  \left(  \lambda
^{\prime}\cdot_ {\left(  \phi\right)  } u\right)  \label{property3}%
\end{equation}
and:%
\begin{equation}
\left(  u +_{\left(  \phi\right)  } v\right)  +_{\left(  \phi\right)  } w=u +_{\left(  \phi\right)  } \left(  v +_{\left(  \phi\right)  } w\right).
\label{property4}%
\end{equation}
Indeed, e.g.:%
\begin{equation}
\lambda  \cdot_ {\left(  \phi\right)  }  \left(  \lambda^{\prime
} \cdot_ {\left(  \phi\right)  }  u\right)  =\phi\left(  \lambda
\phi^{-1}\left(  \lambda^{\prime}  \cdot_ {\left(  \phi\right)  } u\right)  \right)  =\phi\left(  \lambda\lambda^{\prime}\phi^{-1}\left(
u\right)  \right)  =\left(  \lambda\lambda^{\prime}\right)   \cdot_ {\left(  \phi\right)  } u
\end{equation}
which proves (\ref{property3}), and similarly for (\ref{property4}).
$\blacksquare$

To every linear structure there is associated in a canonical way a
\textit{dilation} (or Liouville) field $\Delta$ which is the infinitesimal
generator of dilations (and in fact defines \cite{Vil} the linear structure). Therefore, in the framework of the new
linear structure, it makes sense to consider the mapping:%
\begin{equation}
\Psi:M\times\mathbb{R}\rightarrow M
\end{equation}
via:%
\begin{equation}
\Psi\left(  u,t\right)  =:e^{t}  \cdot_ {\left(  \phi\right)  } u=:u\left(  t\right),
\label{dilation}%
\end{equation}
i.e.:%
\begin{equation}
u\left(  t\right)  =\phi\left(  e^{t}\phi^{-1}(u)\right)  \label{Liouville0}.
\end{equation}

Property \ (\ref{property3}) ensures that:%
\begin{equation}
\Psi\left(  u\left(  t^{\prime}\right)  ,t\right)  =\Psi\left(  u,t+t^{\prime
}\right),
\end{equation}
i.e. that (\ref{dilation}) is indeed a one-parameter group. Then, the
infinitesimal generator of the group is defined as:%
\begin{equation}
\Delta\left(  u\right)  =\left[  \frac{d}{dt}u(t)\right]  _{t=0}=\left[
\frac{d}{dt}\phi\left(  e^{t}\phi^{-1}(u)\right)  \right]  _{t=0}.%
\end{equation}
Explicitly, in components:%
\begin{equation}
\Delta=\Delta^{i}\frac{\partial}{\partial u^{i}} \label{Liouville1}%
\end{equation}
and:%
\begin{equation}
\Delta^{i}=\left[  \frac{\partial\phi^{i}\left(  w\right)  }{\partial w^{j}%
}w^{j}\right]  _{w=\phi^{-1}\left(  u\right)  } .\label{Liouville2}%
\end{equation}
In other words, if we denote by $\Delta_{0}=w^{i}\partial/\partial w^{i}$ the
Liouville field associated with the standard linear structure on $E$:%
\begin{equation}
\Delta=\phi_{\ast}\Delta_{0},
\end{equation}
where $\phi_{\ast}$ denotes, as usual, the push-forward.\\

\textbf{Remark.} If $M=E$ and $\phi$ is a linear (and invertible) map, then
(\ref{Liouville2}) yields: $\Delta^{i}=u^{i}$, i.e.:%
\begin{equation}
\phi_{\ast}\Delta_{0}=\Delta_{0}.
\end{equation}

\bigskip

\textbf{ Examples. \ }

We shall discuss here a couple of examples. Other simple examples are
described in the Appendix.

\begin{itemize}
\item As a first example, consider $T^{\ast}\mathbb{R}$ with coordinates
$\left(  q,p\right)  $ and the standard symplectic form $\omega=dq\wedge dp$.
The linear structure is defined by the dilation (Liouville) field:%
\begin{equation}
\Delta=q\frac{\partial}{\partial q}+p\frac{\partial}{\partial p}%
\end{equation}
and is such that:%
\begin{equation}
i_{\Delta}\omega=qdp-pdq=:2\theta
\end{equation}
and: $\omega=d\theta$. Another relevant structure that can be \ constructed is
the complex structure, that is defined by the $\left(  1,1\right)  $ tensor
field:%
\begin{equation}
J=dp\otimes\frac{\partial}{\partial q}-dq\otimes\frac{\partial}{\partial p},%
\end{equation}
which satisfies $J^{2}=-\mathbb{I}$ (the identity) and, being constant, has a
vanishing Nijenhuis tensor \cite{Nij}: \ $N_{J}=0$.
Notice that:
\begin{equation}
J\circ\omega=g ,\label{metric}%
\end{equation}
where $g$ is the $\left(  2,0\right)  $ tensor:%
\begin{equation}
g=dq\otimes dq+dp\otimes dp,
\end{equation}
i.e. a (Euclidean) metric tensor, and: $g\left( \cdot ,\cdot \right)  =\omega\left(
J \cdot , \cdot \right) $.\\

\textbf{Remark.} In this way we have defined three structures on a cotangent
bundle (actually on the cotangent bundle of a vector space), namely a
symplectic structure, a complex structure and a\ metric tensor. It should be
clear from, e.g., eq.  (\ref{metric}) that, given, say, $g$, we can define in
turn the complex structure as:%
\begin{equation}
J=g\circ\omega^{-1}.%
\end{equation}
In other words, the three structures are not independent: given any two of
them the third one is defined in terms of the previous ones \cite{Sim}.\\

We recall now \cite{Ferr} that with any $\left(
1,1\right)  $ tensor field $S$ with vanishing Nijenhuis tensor one can
associate an antiderivation $d_{S}$ of degree one satisfying\footnote{As a
consequence of the Nijenhuis condition $N_{S}=0$.} $d_{S}^{2}=0$ and that acts
in particular on functions as: $(d_{S}f)\left(  X\right)  =df\left(  S\cdot
X\right)  $ with $X$ any vector field, or:%
\begin{equation}
d_{S}f=\widehat{S}\cdot df,
\end{equation}
with $\widehat{S}$ denoting the action of $S$ on forms. \ Then it is easy to
prove that:%
\begin{equation}
\theta=\frac{1}{2}d_{J}\left(  \frac{1}{2}\left(  p^{2}+q^{2}\right)  \right)
\end{equation}
and hence also:%
\begin{equation}
\omega=\frac{1}{2}dd_{J}\left(  \frac{1}{2}\left(  p^{2}+q^{2}\right)
\right).
\end{equation}

Consider now the dynamics of the $1D$ harmonic oscillator that, in appropriate
units, is described, as is well known, by the vector field:%
\begin{equation}
\Gamma=p\frac{\partial}{\partial q}-q\frac{\partial}{\partial p},%
\end{equation}
which is $\omega$-Hamiltonian: $i_{\Gamma}\omega=dH$ with Hamiltonian:
$H=\left(  q^{2}+p^{2}\right)  /2$. Notice that:%
\begin{equation}
\Gamma=J\left(  \Delta\right).
\end{equation}
Consider the the nonlinear change of coordinates \cite{Vent}: $\left(  q,p\right)
\rightarrow\left(  Q,P\right)  $ with:%
\begin{equation}
\eqalign{Q=q\left(  1+f\left(  H\right)  \right)  \\
 P=p\left(  1+f\left(
H\right)  \right)  .}\label{change}%
\end{equation}
Under very mild assumptions on the function $f\left(  H\right)  $ the mapping
(\ref{change}) will be smooth and invertible with a smooth inverse. One might
assume, e.g., that $f\left(  \cdot \right)  $ be nonnegative and monotonically
increasing for positive argument. \ Then, as:
\begin{equation}
H^{\prime}=:\frac{1}{2}\left(  Q^{2}+P^{2}\right)  =H\left(  1+f\left(
H\right)  \right)  ^{2},%
\end{equation}
one can solve for $H$ and invert the mapping as:%
\begin{equation}
q=\frac{Q}{1+\phi\left(  H^{\prime}\right)  }%
\end{equation}
and similarly for $p$, with: $\phi\left(  H^{\prime}\right)  =:f\left(
H\left(  H^{\prime}\right)  \right)  $. \ 

It is not hard to see that the $1D$ harmonic oscillator, whose dynamics is
now given by:%
\begin{equation}
\Gamma=P\frac{\partial}{\partial Q}-Q\frac{\partial}{\partial P},%
\end{equation}
will be again Hamiltonian with respect to the symplectic form:%
\begin{equation}
\omega^{\prime}=dQ\wedge dP, \label{symp3}%
\end{equation}
with $H^{\prime}$ as Hamiltonian. One can define now a new Liouville field
$\Delta^{\prime}$ via:%
\begin{equation}
i_{\Delta^{\prime}}\omega^{\prime}=QdP-PdQ
\end{equation}
(hence a new linear structure) and a new complex structure:%
\begin{equation}
J^{\prime}=dP\otimes\frac{\partial}{\partial Q}-dQ\otimes\frac{\partial
}{\partial P},%
\end{equation}
and it is clear that:%
\begin{equation}
\Gamma=J\left(  \Delta\right)  =J^{\prime}\left(  \Delta^{\prime}\right) .\label{hamvec}
\end{equation}

In this way we have two different and nonlinearly related linear structures on
$T^{\ast}\mathbb{R\approx R}^{2}$\footnote{The dynamics of the harmonic
oscillator will be compatible with both.}. The $2D$ translation group
$\mathbb{R}^{2}$ is realized then in two different ways, generated by the
vector fields $\partial/\partial q$ and $\partial/\partial p$ in one case and
by $\ \partial/\partial Q$ and $\partial/\partial P$ in the other. \ One
interesting consequence of this is that one obtains two different
ways of defining the Fourier transform\footnote{Remember that the Fourier
transform plays a central r\^{o}le in the implementation of the Weyl
quantization scheme.}. In particular, when considering square-integrable
functions in $L_{2}\left(  \mathbb{R}^{2}\right)  $, functions that are
square-integrable in one coordinate system need not be so in the other, as the
two Lebesgue measures are related by a non-constant Jacobian\footnote{Indeed:
$\partial\left(  Q,P\right)  /\partial(q,p )=\left(  1+f\left(  H\right)  \right)  \left(  1+f\left(  H\right)
+2Hf^{\prime}\left(  H\right)  \right)  \equiv dH^{\prime}/dH$.}.

To be more explicit, let us choose the transformation
\begin{equation}
\eqalign{q = Q  ( 1+  \lambda R^2 )\\
p = P (1+  \lambda R^2 )},
\end{equation}
with $R^2 = P^2 + Q^2$, which can be inverted as
\begin{equation}
\eqalign{Q = q K(r) \\
P = p K(r)},
\end{equation}
where $r ^2 = p^2 + q^2$, and the positive function $K(r)$ is given by the relation  $R = r K(r)$  and satisies the equation $\lambda r^2 K(r)^3 + K(r)-1 =0$.  It is not difficult to check that 
\begin{equation}
\left\vert \begin{array}{c} \frac{\partial}{\partial Q} \\ \frac{\partial}{\partial P}  \end{array} \right\vert = A \left| \begin{array}{c}  \frac{\partial}{\partial q} \\ \frac{\partial}{\partial p}  \end{array} \right\vert
\end{equation}
where
\begin{eqnarray}
A &\equiv & \left\vert \begin{array}{cc} 1+  \lambda (3Q^2+P^2) & 2\lambda PQ \\ 2\lambda PQ & 
1+  \lambda (Q^2+3P^2) \end{array} \right\vert \\
&=& \left\vert \begin{array}{cc} 1+  \lambda K(r)^2 (3q^2+p^2) & 2\lambda K(r)^2 pq \\ 2\lambda K(r)^2 pq & 
1+  \lambda K(r)^2 (q^2+3p^2) \end{array} \right\vert \nonumber 
\end{eqnarray} 
The integral curves in the plane $(q,p)$ of the vector fields $\frac{\partial}{\partial Q}, \frac{\partial}{\partial P}$ are shown in Figure 1.
\begin{figure}
\epsfbox{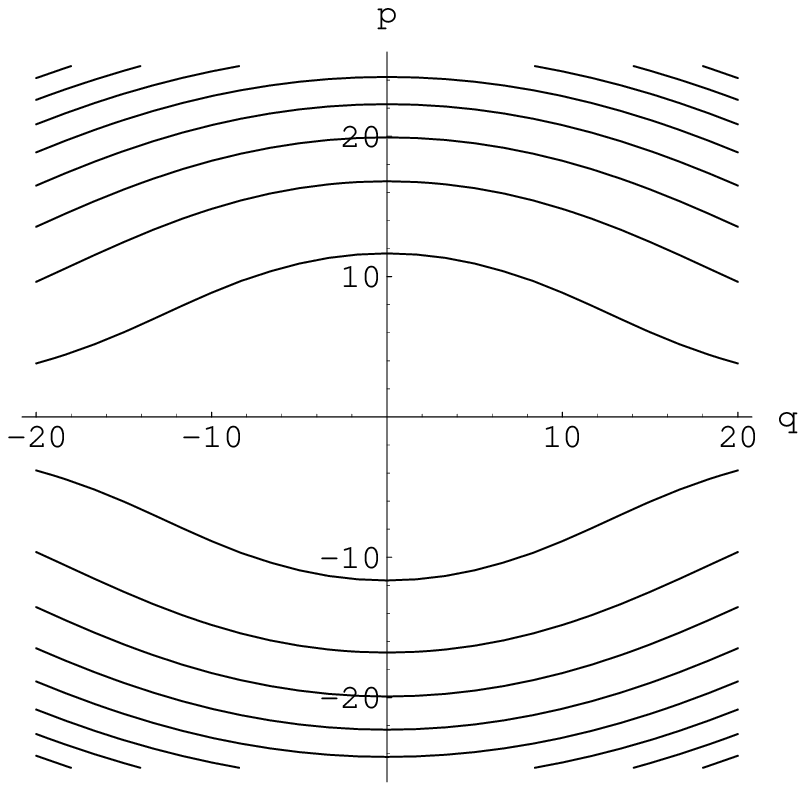}
\epsfbox{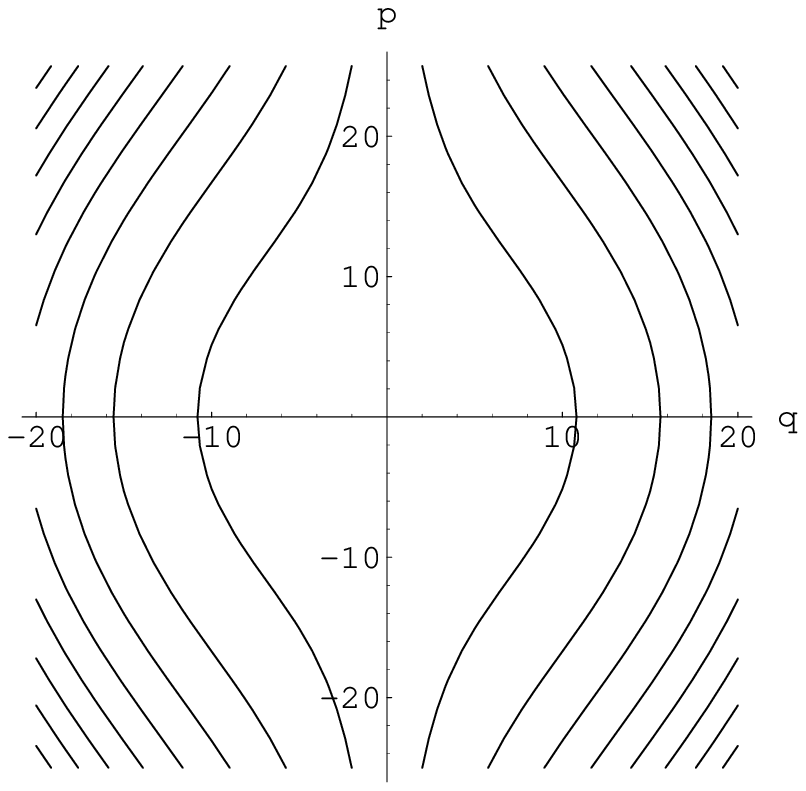}
\caption{\label{a} The integral curves in the plane $(q,p)$ of the vector fields $\frac{\partial}{\partial Q}, \frac{\partial}{\partial P}$. }
\end{figure}

The vector fields $\frac{\partial}{\partial q}, \frac{\partial}{\partial p}$  generate the standard translation group in $\mathbb{R}^2$ associated to the linear structure:
\begin{equation}
\left\vert \begin{array}{c} q\\ p \end{array} \right\vert + \left\vert \begin{array}{c} q^{\prime} \\ p^{\prime}  \end{array} \right\vert = \left\vert \begin{array}{c} q+ q^{\prime} \\ p+ p^{\prime}  \end{array} \right\vert  \label{lin1}
\end{equation}
The alternative linear structure associated to the realization of the translation group by means of the vector fields $\frac{\partial}{\partial Q}, \frac{\partial}{\partial P}$ is instead given by:
\begin{equation}
\eqalign{ \left\vert \begin{array}{c} q\\ p \end{array} \right\vert +_{(K)} \left\vert \begin{array}{c} q^{\prime} \\ p^{\prime}  \end{array} \right\vert =  S(r,r^{\prime})
\left\vert \begin{array}{c} K(r) q + K(r^{\prime}) q^{\prime} \\ K(r) p + K(r^{\prime}) p^{\prime}  \end{array} \right\vert \label{lin2} ,
\\
S(r,r^{\prime}) \equiv  1+\lambda [ K(r)^2 r^2 +K(r^{\prime})^2 
r^{\prime2} + 2 K(r) K(r^{\prime}) (q q^{\prime}+ p p^{\prime})]  .}
\end{equation}

Finally we notice also that 
\begin{equation}
\left\vert \begin{array}{c} dq \\ dp \end{array} \right\vert = A \left| \begin{array}{c} dQ\\ dP \end{array} \right\vert.
\end{equation}
Hence the two symplectic structures $\omega = dq \wedge dp$ and $\omega^{\prime}  = dQ \wedge dP \equiv \omega_K$ with respect to which the vector field (\ref{hamvec}) is Hamiltonian are related by
\begin{equation}
\omega   = D\,  \omega^{\prime} \; \; , \; \; D \equiv \det A
\end{equation}
and define two different Poisson brackets, $\{\cdot,\cdot\}$ and $\{\cdot,\cdot\}_K$ such that
\begin{equation}
\{f,g\}_K =  D \, \{f,g\}  .\label{fg}
\end{equation} 
\end{itemize}

\begin{itemize}
\item A second example that we shall discuss briefly here is borrowed from
Quantum Mechanics, and has to do with a "superposition principle" (better, a
composition rule) for pure states \cite{Man}. Given a Hilbert space $\mathcal{H}$, the
space of pure states is the (complex) projective Hilbert space $\mathcal{PH}$
\ whose "points" are the one-dimensional projectors of the form: $\rho=$
$\left\vert \psi\rangle\langle\psi\right\vert ,$ $\psi\in\mathcal{H},$
$\left\langle \psi|\psi\right\rangle =1$. For finite-dimensional Hilbert
spaces, $\mathcal{H}\approx\mathbb{C}^{n}$ for some $n$ and $\mathcal{PH}%
\approx\mathbb{CP}^{n-1}$. In particular, for a two-level system the
projective Hilbert space can be identified with the two-sphere $\mathbb{S}%
^{2}$ (the Bloch sphere). It is pretty obvious that $\mathcal{PH}$ is not a
vector space. Indeed, e.g., projectors are rank-one operators, while a generic
linear combination of projectors is a rank-two operator. However, one can
define a rule for combining pure states as follows. Select a "fiducial" vector
$|\psi_{0}\rangle$ in the unit sphere in the Hilbert space and the associated
"fiducial" pure state $\rho_{0}=\left\vert \psi_{0}\rangle\langle\psi
_{0}\right\vert $. Given then any two pure states $\rho_{1}$ and $\rho_{2}\,$,
and assuming that they are \textit{not} orthogonal\footnote{In the case of a
two-level system, this will require excluding from the Bloch sphere the point
antipodal to $\rho_{0}$.} to $\rho_{0}$, we can form the linear combination:%
\begin{equation}
|\psi\rangle=c_{1}\rho_{1}|\psi_{0}\rangle+c_{2}\rho_{2}|\psi_{0}%
\rangle; \; c_{1},c_{2}\in\mathbb{C}%
\end{equation}
and then the pure state:%
\begin{equation}
\rho=\frac{\left\vert \psi\rangle\langle\psi\right\vert }{\left\langle
\psi|\psi\right\rangle }=\frac{\left(  c_{1}\rho_{1}|\psi_{0}\rangle+c_{2}%
\rho_{2}|\psi_{0}\rangle\right)  \left(  \overline{c}_{1}\langle\psi_{0}%
|\rho_{1}+\overline{c}_{2}\langle\psi_{0}|\rho_{2}\right)  }{\left\Vert
c_{1}\rho_{1}|\psi_{0}\rangle+c_{2}\rho_{2}|\psi_{0}\rangle\right\Vert ^{2}}.%
\end{equation}

\end{itemize}

This procedure will define a composition rule (not a linear superposition, of
course) for pure states. It will depend on the choice of the fiducial vector
(i.e. of $\rho_{0}$ \textit{and} of a phase factor), but also on the linear
structure of the underlying vector space. Even if $|\psi_{0}\rangle$ is kept
fixed, alternative linear structures in $\mathcal{H}$ (or $\mathbb{C}^{n}$ for
a system with a finite number of levels) will define therefore alternative
composition rules on the projective Hilbert space. \\

\textbf{Remark. }The case of a two-level system (when $\mathcal{PH}%
\approx\mathbb{S}^{2}$) offers another interesting possibility.
Indeed\footnote{See the last example discussed in the Appendix.} one can
induce a linear structure on the punctured sphere via stereographic projection
from $\mathbb{R}^{2}$. In this way, projecting from the "excluded" point (the
projector orthogonal to $\rho_{0}$) on can \ induce a truly \textit{linear}
structure on the space of pure states of a two-level system, provided one
excludes a single (pure) state.

\section{Linear Structures Associated to Regular Lagrangians}

A regular Lagrangian $\mathcal{L}$ will define the symplectic structure on
$TQ$:%
\begin{equation}
\omega_{\mathcal{L}}=d\theta_{\mathcal{L}}=d\left(  \frac{\partial\mathcal{L}%
}{\partial u^{i}}\right)  \wedge dq^{i};\; \theta_{\mathcal{L}}=\left(
\frac{\partial\mathcal{L}}{\partial u^{i}}\right)  dq^{i}.%
\end{equation}
We look now \cite{Mar}  for Hamiltonian vector
fields $X_{j},Y^{j}$ such that:%
\begin{equation}
i_{X_{j}}\omega_{\mathcal{L}}=-d\left(  \frac{\partial\mathcal{L}}{\partial
u^{j}}\right)  ,\; i_{Y^{j}}\omega_{\mathcal{L}}=dq^{j}%
\end{equation}
which implies, of course:%
\begin{equation}
L_{X_{j}}\omega_{\mathcal{L}}=L_{Y^{j}}\omega_{\mathcal{L}}=0.
\end{equation}
Explicitly:%
\begin{equation}
i_{X_{j}}\omega_{\mathcal{L}}=\left(  L_{X_{j}}\frac{\partial\mathcal{L}%
}{\partial u^{i}}\right)  dq^{i}-d\left(  \frac{\partial\mathcal{L}}{\partial
u^{i}}\right)  \left(  L_{X_{j}}q^{i}\right)
\end{equation}
and this implies:%
\begin{equation}
L_{X_{j}}q^{i}=\delta_{j}^{i},\; L_{X_{j}}\frac{\partial\mathcal{L}%
}{\partial u^{i}}=0 \label{condition1}.%
\end{equation}
Similarly:%
\begin{equation}
i_{Y^{j}}\omega_{\mathcal{L}}=\left(  L_{Y^{j}}\frac{\partial
\mathcal{L}}{\partial u^{i}}\right)  dq^{i}-d\left(  \frac{\partial
\mathcal{L}}{\partial u^{i}}\right)  \left(  L_{Y^{j}}q^{i}\right)
\end{equation}
and this implies in turn:%
\begin{equation}
L_{Y^{j}}q^{i}=0,\;L_{Y^{j}}\frac{\partial\mathcal{L}}{\partial u^{i}%
}=\delta_{i}^{j} \label{condition2}.%
\end{equation}
Using then the identity:%
\begin{equation}
i_{\left[  Z,W\right]  }=L_{Z}\circ i_{W}-i_{W}\circ L_{Z},%
\end{equation}
we obtain, whenever both ~$Z$ and $W$ are Hamiltonian ($i_{Z}\omega
_{\mathcal{L}}=dg_{Z}$ and similarly for $W$):%
\begin{equation}
i_{\left[  Z,W\right]  }\omega_{\mathcal{L}}=d\left(  L_{Z}g_{W}\right).
\end{equation}
Taking now: $(Z,W)=(X_{i},X_{j}),(X_{i},Y^{j})$ or $(Y^{i},Y^{j})$, the Lie
derivative of the Hamitonian of every field with respect to any other field is either
zero or a constant (actually unity). Therefore:%
\begin{equation}
i_{\left[  Z,W\right]  }\omega_{\mathcal{L}}=0\; \mbox{\rm{whenever}} \; \left[  Z,W\right]  =\left[  X_{i},X_{j}\right]  ,\left[  X_{i},Y^{j}\right]
,\left[  Y^{i},Y^{j}\right],
\end{equation}
which proves that:
\begin{equation}
\left[  X_{i},X_{j}\right]  =\left[  X_{i},Y^{j}\right]  =\left[  Y^{i}
,Y^{j}\right]  =0 .
\end{equation}
This defines an infinitesimal action of an Abelian Lie group on $TQ$. If this
integrates to an action of the group $\mathbb{R}^{2n}$ ($\dim Q=n$) that is
free and transitive, this will define a new vector space structure on $TQ$
 that is "adapted" to the Lagrangian two-form $\omega_{\mathcal{L}}$.

Spelling now explicitly eq.ns (\ref{condition1}) and (\ref{condition2}) we
find that $X_{j}$ and $Y^{j}$ must be of the form:%
\begin{equation}
X_{j}=\frac{\partial}{\partial q^{j}}+\left(  X_{j}\right)  ^{k}\frac
{\partial}{\partial u^{k}},\; Y^{j}=\left(  Y^{j}\right)  ^{k}%
\frac{\partial}{\partial u^{k}};\; \left(  X_{j}\right)  ^{k},\left(
Y^{j}\right)  ^{k}\in\mathcal{F}\left(  TQ\right)
\end{equation}
and that:%
\begin{equation}
L_{X_{j}}\frac{\partial\mathcal{L}}{\partial u^{i}}=0\Rightarrow\frac
{\partial^{2}\mathcal{L}}{\partial u^{i}\partial q^{j}}+\left(  X_{j}\right)
^{k}\frac{\partial^{2}\mathcal{L}}{\partial u^{i}\partial u^{k}}=0
\label{condition3}%
\end{equation}
and:%
\begin{equation}
L_{Y^{j}}\frac{\partial\mathcal{L}}{\partial u^{i}}=\delta_{j}^{i}%
\Rightarrow\left(  Y^{j}\right)  ^{k}\frac{\partial^{2}\mathcal{L}}{\partial
u^{i}\partial u^{k}}=\delta_{i}^{j} \label{condition4}.%
\end{equation}
Therefore, the Hessian being not singular by assumption, \ $\left(
Y^{j}\right)  ^{k}$ is the inverse of the Hessian matrix, while $\left(
X_{j}\right)  ^{k}$ can be obtained algebraically from Eqn.(\ref{condition3}).

Defining dual forms $\left(  \alpha^{i},\beta_{i}\right)  $ via:%
\begin{equation}
\alpha^{i}\left(  X_{j}\right)  =\delta_{j}^{i},\;\alpha^{i}\left(
Y^{j}\right)  =0
\end{equation}
and similarly:%
\begin{equation}
\beta_{i}\left(  Y^{j}\right)  =\delta_{i}^{j},\;\beta_{i}\left(
X_{j}\right)  =0.
\end{equation}
Testing then the identity:%
\begin{equation}
d\theta\left(  Z,W\right)  =L_{Z}\left(  \theta(W\right)  -L_{W}\left(
\theta\left(  Z\right)  \right)  -\theta\left(  \left[  Z,W\right]  \right)
\end{equation}
on the pairs $(Z,W)=(X_{i},X_{j}),(X_{i},Y^{j}),\left(  Y^{i},Y^{j}\right)  $,
one proves immediately that the dual forms are all closed.$\blacksquare$

 Moreover, it is also immediate to see that:%
\begin{equation}
\alpha^{i}=dq^{i}%
\end{equation}
and:%
\begin{equation}
\beta_{i}=d\left(  \frac{\partial\mathcal{L}}{\partial u^{i}}\right)
\end{equation}
and that the symplectic form can be written as:%
\begin{equation}
\omega_{\mathcal{L}}=\beta_{i}\wedge\alpha^{i}.%
\end{equation}

Basically, what this means is that, to the extent that the definition of
vector fields and dual forms is global, we have found in this way a global
Darboux chart.

\setcounter{footnote}{1}
\subsection{Examples of Adapted Linear Structures for Lagrangian Systems}

\begin{itemize}
\item For the "standard" Lagrangian:%
\begin{equation}
\mathcal{L=}\frac{1}{2}\delta_{ij}u^{i}u^{j}-U\left(  q\right)
\end{equation}
the solution is of course the standard one, i.e.:%
\begin{equation}
X_{j}=\frac{\partial}{\partial q^{j}},Y^{j}=\delta^{jk}\frac{\partial
}{\partial u^{k}}.%
\end{equation}

\item A particle in a (time-independent) magnetic field $\overrightarrow
{B}=\nabla\times\overrightarrow{A}$. \ The corresponding second-order vector
field is given by ($e=m=c=1$):%
\begin{equation}
\Gamma=u^{i}\frac{\partial}{\partial q^{i}}+\delta^{is}\epsilon_{ijk}%
u^{j}B^{k}\frac{\partial}{\partial u^{s}} \label{vectorfield}%
\end{equation}
and the equations of motion are:%
\begin{equation}
\frac{dq^{i}}{dt}=u^{i},\;\frac{du^{i}}{dt}=\delta^{ir}\epsilon
_{rjk}u^{j}B^{k}\;,\;i=1,2,3 .\label{equmotion}%
\end{equation}
The Lagrangian is given in turn by :%
\begin{equation}
\mathcal{L}=\frac{1}{2}\delta_{ij}u^{i}u^{j}+u^{i}A_{i}.%
\end{equation}
Hence:%
\begin{equation}
\theta_{\mathcal{L}}=\frac{\partial\mathcal{L}}{\partial u^{i}}dq^{i}=\left(
\delta_{ij}u^{j}+A_{i}\right)  dq^{i} \label{Cartan}%
\end{equation}
and the symplectic form is\footnote{As: $\theta_{\mathcal{L}}=\theta
_{\mathcal{L}}^{\left(  0\right)  }+A$, $\theta_{\mathcal{L}}^{\left(
0\right)  }=\delta_{ij}u^{j}dq^{i},A=A_{i}dq^{i}$, then: $dA=:B=\frac{1}%
{2}\varepsilon_{ijk}B^{i}dq^{j}\wedge dq^{k}$, and: $\omega_{\mathcal{L}%
}=\omega_{0}-B$.}:%
\begin{equation}
\omega_{\mathcal{L}}=-d\theta_{\mathcal{L}}=\delta_{ij}dq^{i}\wedge
du^{j}-\frac{1}{2}\varepsilon_{ijk}B^{i}dq^{j}\wedge dq^{k}.%
\end{equation}
The field \ $\Gamma$ satisfies:%
\begin{equation}
i_{\Gamma}\omega_{\mathcal{L}}=dH,
\end{equation}
with the Hamiltonian:%
\begin{equation}
H=\frac{1}{2}\delta_{ij}u^{i}u^{j}.%
\end{equation}
Now it is easy to see that:%
\begin{equation}
X_{j}=\frac{\partial}{\partial q^{j}}-\delta^{ik}\frac{\partial A_{k}%
}{\partial q^{j}}\frac{\partial}{\partial u^{i}},
\end{equation}
while:%
\begin{equation}
Y^{j}=\delta^{jk}\frac{\partial}{\partial u^{k}}.%
\end{equation}
Dual forms\ $\alpha^{i},\beta_{i},i=1,...,n=\dim Q$ are defined as:%
\begin{equation}
\eqalign{\left\langle \alpha^{i}|X_{j}\right\rangle =\delta_{j}^{i},\; \left\langle \alpha^{i}|Y^{j}\right\rangle =0,
\\
\left\langle \beta_{i}|X_{j}\right\rangle =0,\;\left\langle \beta
_{i}|Y^{j}\right\rangle =\delta_{i}^{j} ,}
\end{equation}
and one finds easily:%
\begin{equation}
\eqalign{ \alpha^{i}=dq^{i},\\
\beta_{i}=\delta_{ij}dU^{j},\;U^{j}=:u^{j}+\delta^{jk}A_{k}.}
\end{equation}
Notice that in this way the Cartan form (\ref{Cartan}) acquires the form:%
\begin{equation}
\theta_{\mathcal{L}}=\pi_{i}dq^{i},%
\end{equation}
where:%
\begin{equation}
\pi_{i}=\delta_{ij}u^{j}+A_{i},%
\end{equation}
and the symplectic form becomes:%
\begin{equation}
\omega_{\mathcal{L}}=dq^{i}\wedge d\pi_{i} \label{symp2}.%
\end{equation}
It appears therefore that the mapping:%
\begin{equation}
\phi:\left(  q,u\right)  \rightarrow\left(  Q,U\right)  ,\label{mapping1}%
\end{equation}
with:%
\begin{equation}%
\eqalign{
Q^{i}=q^{i}\\
U^{i}=u^{i}+\delta^{ik}A_{k}, }
\label{mapping2}%
\end{equation}
(hence: $\pi_{i}=\delta_{ij}U^{j}$) provides us with a symplectomorphism that
reduces $\omega_{\mathcal{L}}$ to the canonical form, i.e. that the chart
$\left(  Q,U\right)  $ is a Darboux chart "adapted" to the vector potential
$\overrightarrow{A}$.

The mapping (\ref{mapping2}) is clearly invertible, and:%
\begin{equation}
\frac{\partial q^{i}}{\partial Q^{j}}=\delta_{j}^{i},\;\frac{\partial
q^{i}}{\partial U^{j}}=0,
\end{equation}
while:%
\begin{equation}
\frac{\partial u^{i}}{\partial U^{j}}=\delta_{j}^{i},\;\frac{\partial
u^{i}}{\partial Q^{j}}=-\delta^{ik}\frac{\partial A_{k}}{\partial Q^{j}},%
\end{equation}
$A_{k}\left(  q\right)  \equiv A_{k}\left(  Q\right)  $. But then:%
\begin{equation}
X_{j}=\frac{\partial}{\partial Q^{j}},\;Y^{j}=\delta^{jk}\frac{\partial
}{\partial U^{k}},%
\end{equation}
as well as:%
\begin{equation}
\alpha^{i}=dQ^{i},\;\beta_{i}=d\pi_{i}=\delta_{ij}dU^{j}.%
\end{equation}

The push-forward of the Liouville field: $\Delta_{0}=q^{i}\partial/\partial
q^{i}+u^{i}\partial/\partial u^{i}$ will be then:%
\begin{equation}
\Delta=\phi_{\ast}\Delta_{0}=Q^{i}\frac{\partial}{\partial Q^{i}}+\left[
U^{i}+\delta^{ik}\left(  Q^{j}\frac{\partial A_{k}}{\partial Q^{j}}%
-A_{k}\right)  \right]  \frac{\partial}{\partial U^{i}} \label{newlinear}.%
\end{equation}

\end{itemize}

\textbf{Remarks.}

\begin{enumerate}
\item As remarked previously: $\phi_{\ast}\Delta_{0}=\Delta_{0}$ whenever the
vector potential is homogeneous of degree one in the coordinates (constant
magnetic field) an hence the mapping (\ref{mapping2}) is linear.

\item For an arbitrary vector potential the linear structure $\Delta$ depends
on the gauge choice. This is a consequence of the mapping (\ref{mapping2})
being also gauge-dependent, which means in turn that every choice of gauge
will define a \textit{different} linear structure. The symplectic form
(\ref{symp2}) will be however gauge-independent.

\item Denoting collectively the old and new coordinates as $\left(
q,u\right)  $ and $\left(  Q,U\right)  $ respectively, eq. (\ref{mapping2})
defines a mapping:%
\begin{equation}
\left(  q,u\right)  \stackrel{\phi}{\rightarrow}\left(  Q,U\right).
\end{equation}
It is then a straightforward application of the definitions (\ref{property1})
and (\ref{property2}) to show that the rules of addition and multiplication by
a constant become, in this specific case:%
\begin{equation}
\fl \left(  Q,U\right)  +_{\left(  \phi\right)  } \left(  Q^{\prime
},U^{\prime}\right)  =\left(  Q+Q^{\prime},U+U^{\prime}+\left[  A\left(
Q+Q^{\prime}\right)  -\left(  A(Q)+A(Q^{\prime}\right)  )\right]  \right),
\label{sum2}%
\end{equation}
and:%
\begin{equation}
\lambda \cdot_{\left(  \phi\right)  } \left(  Q,U\right)  =\left(
\lambda Q,\lambda U+\left[  A\left(  \lambda Q\right)  -\lambda A\left(
Q\right)  \right]  \right)  . \label{product2}%
\end{equation}
In particular, with $\lambda=e^{t}$, the infinitesimal version of
(\ref{product2}) yields precisely the infinitesimal generator (\ref{newlinear}%
) and, if the vector potential is, as in the case of a constant magnetic
field, homogeneous of degree one in the coordinates, all the terms in square
brackets in eq.ns (\ref{sum2}) and (\ref{product2}) vanish identically, as expected.

\item Notice that the origin of the new linear structure is given by:
$\phi\left(  0,0\right)  =\left(  0,A\left(  0\right)  \right)  $ \ and,
correctly: $0 \cdot_{\left(  \phi\right)  }\left(  Q,U\right)
=\left(  0,A\left(  0\right)  \right)$  $\forall\left(  Q,U\right)  $ as well
as: $\lambda \cdot_{\left(  \phi\right)  }\left(  0,A(0)\right)
=\left(  0,A\left(  0\right)  \right)$  $\forall\lambda$.  Moreover: $\left(
Q,U\right)  +\left(  0,A\left(  0\right)  \right)  =\left(  Q,U\right)  $ 
$\forall\left(  Q,U\right)  $. Finally, the difference between\ any two points
$\left(  Q,U\right)  $ and $\left(  Q^{\prime},U^{\prime}\right)  $ must be
understood as:%
\begin{equation}
\left(  Q,U\right)  -_{\left(  \phi\right)  } \left(  Q^{\prime
},U^{\prime}\right)  =:\left(  Q,U\right)  +_{\left(  \phi\right)  } \left(  \left(  -1\right)  \cdot_{\left(  \phi\right)  } \left(
Q^{\prime},U^{\prime}\right)  \right)
\end{equation}
and, because of: $\left(  -1\right)  \cdot_{\left(  \phi\right)  } \left(
Q^{\prime},U^{\prime}\right)  =\left(  -Q^{\prime},-U^{\prime}+A\left(
Q^{\prime}\right)  +A\left(  -Q^{\prime}\right)  \right)  $,
we finally get:%
\begin{equation}
\fl \left(  Q,U\right)  -_{\left(  \phi\right)  }\left(  Q^{\prime
},U^{\prime}\right)  = \left(  Q-Q^{\prime},U-U^{\prime}+A(Q-Q^{\prime}\right)
+A\left(  Q^{\prime}\right)  -A\left(  Q\right)).
\end{equation}
Again, if $Q^{\prime}=Q,U^{\prime}=U,$ $\left(  Q,U\right)  -_{\left(  \phi\right)  }\left(  Q,U\right)  =\left(  0,A\left(  0\right)  \right)  $.
\end{enumerate}

If we work with the standard Euclidean metric, there is actually no need to
distinguish between uppercase and lowercase indices ($Q_{i}=:\delta_{ij}%
Q^{j}=Q^{i}$ etc.). Then, the push-forward of the dynamical vector field is:%
\begin{equation}
\widetilde{\Gamma}=\phi_{\ast}\Gamma=\left(  U^{i}-A^{i}\right)
\frac{\partial}{\partial Q^{i}}+\left(  U^{k}-A^{k}\right)  \frac{\partial
A_{k}}{\partial Q^{i}}\frac{\partial}{\partial U^{i}}%
\end{equation}
and is Hamiltonian with respect to the symplectic form (\ref{symp2}) with the
Hamitonian:%
\begin{equation}
\widetilde{H}=\phi^{\ast}H=\frac{1}{2}\delta_{ij}\left(  U^{i}-A^{i}\right)
\left(  U^{j}-A^{j}\right)  \label{Ham}.%
\end{equation}

\begin{itemize}
\item In particular, for a constant magnetic field $B=\left(  0,0,B\right)$ with, e.g., the vector potential in the symmetric gauge:%

\begin{equation}
\overrightarrow{A}=\frac{B}{2}\left(  -q^{2},q^{1},0\right)  =\frac{1}%
{2}\overrightarrow{B}\times\overrightarrow{r},\;\overrightarrow
{B}=B\widehat{k}\Rightarrow\;A_{i}=\frac{1}{2}\varepsilon_{ijk}%
B^{j}q^{k},%
\end{equation}

\begin{equation}
X_{1}=\frac{\partial}{\partial q^{1}}-\frac{B}{2}\frac{\partial}{\partial
u^{2}},\;X_{2}=\frac{\partial}{\partial q^{2}}+\frac{B}{2}\frac
{\partial}{\partial u^{1}},\;X_{3}=\frac{\partial}{\partial q^{3}},%
\end{equation}

\begin{equation}
\alpha^{i}=dq^{i}%
\end{equation}
and:%
\begin{equation}
\beta_{1}=du^{1}-\frac{B}{2}dq^{2},\;\beta_{2}=du^{2}+\frac{B}{2}%
dq^{1},\;\beta_{3}=du^{3},%
\end{equation}
while (see above) $\Delta=\Delta_{0}$, as expected.

According to eq.ns (\ref{mapping2}) and (\ref{equmotion}), the equations of
motion in the new coordinates are given by:%
\begin{equation}
\frac{d}{dt}\left\vert
\begin{array}
[c]{c}%
Q^{1}\\
Q^{2}\\
U^{1}\\
U^{2}%
\end{array}
\right\vert =\mathbb{G}\left\vert
\begin{array}
[c]{c}%
Q^{1}\\
Q^{2}\\
U^{1}\\
U^{2}%
\end{array}
\right\vert,
\end{equation}
where:%

\begin{equation}
\mathbb{G}=\left\Vert G^{i}\;_{j}\right\Vert =\left\vert
\begin{array}
[c]{cccc}%
0 & B/2 & 1 & 0\\
-B/2 & 0 & 0 & 1\\
-B^{2}/4 & 0 & 0 & B/2\\
0 & -B^{2}/4 & -B/2 & 0
\end{array}
\right\vert.
\end{equation}

\bigskip 
In other words (cfr. Eqn.(\ref{mapping1})):%
\begin{equation}%
\eqalign{
\phi_{\ast}\Gamma&=\left(  U^{1}+\frac{B}{2}Q^{2}\right)  \frac{\partial}{\partial Q^{1}}+
 \left(  U^{2}-\frac{B}{2}Q^{1}\right)  \frac{\partial
}{\partial Q^{2}}\\
& +\frac{B}{2}\left(  U^{2}-\frac{B}{2}Q^{1}\right)  \frac{\partial}{\partial
U^{1}}-\frac{B}{2}\left(  U^{1}+\frac{B}{2}Q^{2}\right)  \frac{\partial}{\partial U^{2} } .}
\end{equation}

As the transformation (\ref{mapping2}) is not a
point-transformation\footnote{It is the identity on the base and acts only
along the fibers.}, it comes to no surprise that the transformed vector
field is no more a second-order field in the new coordinates. \ However,
$\phi_{\ast}\Gamma$ is still Hamiltonian with respect to the symplectic form
$\phi^{\ast}\omega_{\mathcal{L}}=dQ^{i}\wedge dU_{i}$ \ with \ Hamiltonian:%
\begin{equation}
\phi^{\ast}H=\frac{1}{2}\delta_{ij}(U^{i}-\delta^{ik}A_{k})(U^{j}-\delta
^{jk}A_{k}).
\end{equation}

\end{itemize}

Spelled out explicitly, the equations of motion in the $\left(  Q,U\right)  $
coordinates are:%
\begin{equation}%
\eqalign{
\frac{dQ^{1}}{dt}=U^{1}+\frac{B}{2}Q^{2}&,\; \frac{dQ^{2}}{dt}%
=U^{2}-\frac{B}{2}Q^{1},\\
\frac{dU^{1}}{dt}=\frac{B}{2}\left(  U^{2}-\frac{B}{2}Q^{1}\right) & ,\; \frac{dU^{2}}{dt}=-\frac{B}{2}\left(  U^{1}+\frac{B}{2}Q^{2}\right).
}\label{equ3}%
\end{equation}
Hence: 
\begin{equation}
\eqalign{
\frac{dU^{1}}{dt}=\frac{B}{2}  \frac{dQ^{2}}{dt},\\
\frac{dU^{2}}{dt}=-\frac{B}{2}  \frac{dQ^{1}}{dt}. }
\end{equation}

Therefore:%
\begin{equation}
\chi_{1}=:U^{1}-\frac{B}{2}Q^{2}\mbox{ and: } \chi_{2}=U^{2}%
+\frac{B}{2}Q^{1}\; \label{chi}%
\end{equation}
are constants of the motion\footnote{In fact they are proportional to the
coordinates of the center of the Larmor orbit \cite{Mor}. See also eq.ns (\ref{equ5}) and (\ref{equ6}) below.}, and
this allows an easy integration of the equations of motion. Indeed, using
(\ref{chi}) one finds at once:%
\begin{equation}%
\eqalign{
\frac{dQ^{1}}{dt}=\chi_{1}+BQ^{2},\\
\frac{dQ^{2}}{dt}=\chi_{2}-BQ^{1}. }
\label{equ4}%
\end{equation}
and, setting:%
\begin{equation}
Q^{1}\left(  t\right)  =\frac{\chi_{2}}{B}+\widetilde{Q}^{1}\left(  t\right)
,\;Q^{2}\left(  t\right)  =-\frac{\chi_{1}}{B}+\widetilde{Q}%
^{2}\left(  t\right) , \label{equ5}%
\end{equation}
the $\widetilde{Q}^{i}$'s \ obey the equations:%
\begin{equation}
\frac{d\widetilde{Q}^{1}}{dt}=B\widetilde{Q}^{2},\;\frac{d\widetilde
{Q}^{2}}{dt}=-B\widetilde{Q}^{1}\Rightarrow \frac{d^{2}\widetilde
{Q}^{i}}{dt^{2}}+B^{2}\widetilde{Q}^{i}=0,\;i=1,2 .\label{equ6}%
\end{equation}
These integrate easily and, using again eq.ns (\ref{equ3}), the final result
is:%
\begin{equation}
\left\vert
\begin{array}
[c]{c}%
Q^{1}\left(  t\right) \\
Q^{2}\left(  t\right) \\
U^{1}\left(  t\right) \\
U^{2}\left(  t\right)
\end{array}
\right\vert =\mathbb{F}\left(  t\right)  \left\vert
\begin{array}
[c]{c}%
Q^{1}\\
Q^{2}\\
U^{1}\\
U^{2}%
\end{array},
\right\vert
\end{equation}
where: $Q^{1}=Q^{1}\left(  0\right)  $ etc., and \ $\mathbb{F}\left(
t\right)  =:\exp\left\{  t\mathbb{G}\right\}  $ is given explicitly by:%
\begin{equation}
\mathbb{F}\left(  t\right)  =\left\vert
\begin{array}
[c]{cccc}%
\frac{  1+\cos\left(  Bt\right) }{2} & \frac{\sin\left(  Bt\right)}{2} &
\frac{\sin\left(  Bt\right) }{B} & \frac{ 1-\cos\left(  Bt\right)  }{B}\\
-\frac{\sin\left(  Bt\right)}{2} &\frac{  1+\cos\left(  Bt\right) }{2} &
\frac{  \cos(Bt)  -1}{B }& \frac{\sin\left(  Bt\right) }{B}\\
-\frac{B\sin\left(  Bt\right)}{4} & \frac{ B  \left(  \cos\left(  Bt \right)  -1\right)}{4} &
\frac{ 1+\cos\left(  Bt\right) }{2} & \frac{\sin\left(  Bt\right) }{2}\\
\frac{B\left(  1-\cos\left(  Bt\right)  \right)  }{4 }& -\frac{B\sin\left(  Bt\right) }{4} &
-\frac{\sin\left(  Bt\right) }{2 }& \frac{  1+\cos\left(  Bt\right)  }{  2}
\end{array}
\right\vert.
\end{equation}

\textbf{Checks.}

That $\mathbb{F}\left(  0\right)  =\mathbb{I}$ can be checked by inspection. Moreover:%

\begin{equation}
\fl \frac{d\mathbb{F}}{dt}=\left\vert
\begin{array}
[c]{cccc}%
-\left(  B/2\right)  \sin\left(  Bt\right)  & \left(  B/2\right)  \cos\left(
Bt\right)  & \cos\left(  Bt\right)  & \sin\left(  Bt\right) \\
-\left(  B/2\right)  \cos\left(  Bt\right)  & -\left(  B/2\right)  \sin\left(
Bt\right)  & -\sin\left(  Bt\right)  & \cos\left(  Bt\right) \\
-\left(  B^{2}/4\right)  \cos\left(  Bt\right)  & -\left(  B^{2}/4\right)
\sin\left(  Bt\right)  & -\left(  B/2\right)  \sin\left(  Bt\right)  & \left(
B/2\right)  \cos\left(  Bt\right) \\
\left(  B^{2}/4\right)  \sin\left(  Bt\right)  & -\left(  B^{2}/4\right)
\cos\left(  Bt\right)  & -\left(  B/2\right)  \cos\left(  Bt\right)  &
-\left(  B/2\right)  \sin\left(  Bt\right)
\end{array}
\right\vert.
\end{equation}
Hence: $\left(  d\mathbb{F}/dt\right)  _{t=0}=\mathbb{G}$ as it should be.
That $\mathbb{F}^{-1}\left(  d\mathbb{F}/dt\right)  \equiv_{\left(
t\right)  } \mathbb{G}$ should instead be checked numerically.

Also one should check that:%
\begin{equation}
\widetilde{\mathbb{F}}\cdot\Omega_{D}\cdot\mathbb{F}=\Omega_{D}.%
\end{equation}
This is equivalent to
\begin{equation}
\widetilde{\mathbb{G}}\cdot\Omega_{D}+\Omega_{D}\cdot\mathbb{G}=0,
\label{transpose}%
\end{equation}
where:%
\begin{equation}
\Omega_{D}=\left\vert
\begin{array}
[c]{cc}%
\mathbf{0}_{2\times2} & \mathbb{I}_{2\times2}\\
-\mathbb{I}_{2\times2} & \mathbf{0}_{2\times2}%
\end{array}
\right\vert
\end{equation}
and this is easily checked.

\section{Weyl Systems, Quantization and the Von Neumann Uniqueness
Theorem}

We recall here briefly how Weyl systems are defined and how one can implement
the Weyl-Wigner-von Neumann quantization programme. Let \ $\left(
E,\omega\right)  $ be a symplectic vector space with $\omega$ a constant
symplectic form. A \textit{Weyl system} is a strongly continuous map:
$\mathcal{W}:E\rightarrow\mathcal{U}\left(  \mathcal{H}\right)  $ from $\ E$
\ to the set of unitary operators on some Hilbert space $\mathcal{H}$
\ satisfying \ (we set here $\hbar=1$ for simplicity):%
\begin{equation}
\mathcal{W}\left(  e_{1}\right)  \mathcal{W}\left(  e_{2}\right)  =e^{\frac
{i}{2}\omega\left(  e_{1},e_{2}\right)  }\mathcal{W}\left(  e_{1}%
+e_{2}\right)  ;\;e_{1},e_{2}\in\mathcal{H}%
\end{equation}
or\footnote{This is also called the "Weyl form" of the commutation
relations.}:%
\begin{equation}
\mathcal{W}\left(  e_{1}\right)  \mathcal{W}\left(  e_{2}\right)
=e^{i\omega\left(  e_{1},e_{2}\right)  }\mathcal{W}\left(  e_{2}\right)
\mathcal{W}\left(  e_{1}\right).
\end{equation}

It is clear that operators associated with vectors on a Lagrangian subspace
will commute pairwise and can then be diagonalized simultaneously. Von
Neumann's theorem states then that: $a)$ Weyl systems do exist for any
finite-dimensional symplectic vector space and: $b)$ the Hilbert space
$\mathcal{H}$ can be realized as the space of square-integrable complex
functions on a Lagrangian subspace $L\subset E$ \ with the
translationally-invariant Lebesgue measure. Decomposing then $E$ as $L\oplus
L^{\ast}$, one can define $\mathcal{U}=:\mathcal{W}|_{L^{\ast}}$ and
$\mathcal{V}=:\mathcal{W}|_{L}$ and realize their action on $\mathcal{H}%
=L^{2}\left(  L,d^{n}x\right)  $ \ ($\dim E=2n$) as:
\begin{equation}%
\eqalign{
\left(  \mathcal{V}\left(  x\right)  \psi\right)  \left(  y\right)
=\psi\left(  x+y\right)  \\
\left(  \mathcal{U}\left(  \alpha\right)  \psi\right)  \left(  y\right)
=e^{i\alpha\left(  y\right)  }\psi\left(  y\right)  \\
x,y\in L,\; \alpha\in L^{\ast} .}
\end{equation}

As a consequence of the strong continuity of the mapping $\mathcal{W}$ one can
write, using Stone's theorem \cite{Reed}:%
\begin{equation}
\mathcal{W}\left(  e\right)  =\exp\left\{  i\mathcal{R}\left(  e\right)
\right\}  \;\forall e\in E,
\end{equation}
where $\mathcal{R}\left(  e\right)  $, which depends linearly on $e$, is the
self-adjoint generator of the one-parameter unitary group $\mathcal{W}\left(
te\right)  ,t\in\mathbb{R}$.

If $\left\{  \mathbb{T}\left(  t\right)  \right\}  _{t\in\mathbb{R}}$ is a
one-parameter group of symplectomorphisms ($\mathbb{T}\left(  t\right)
\mathbb{T}\left(  t^{\prime}\right)  =\mathbb{T}\left(  t+t^{\prime}\right)$
 $\forall t,t^{\prime}$ and: $\mathbb{T}^{t}\left(  t\right)  \omega
\mathbb{T}\left(  t\right)  =\omega$ $\forall t$), then we can define:%
\begin{equation}
\mathcal{W}_{t}\left(  e\right)  =:\mathcal{W}\left(  \mathbb{T}\left(
t\right)  e\right).
\end{equation}
This being an automorphism of the unitary group will be inner and will be
therefore represented as a conjugation with a unitary transformation belonging
to a one-parameter unitary group associated with the group $\left\{
\mathbb{T}\left(  t\right)  \right\}  $. If $\mathbb{T}\left(  t\right)  $
represents the dynamical evolution associated with a linear vector field, then
we can write:%
\begin{equation}
\mathcal{W}_{t}\left(  e\right)  =e^{it\widehat{H}}\mathcal{W}\left(
e\right)  e^{-it\widehat{H}}%
\end{equation}
and $\widehat{H}$ will be (again in units $\hbar=1$) the quantum Hamiltonian
of the system.

Uniqueness part of Von Neumann's theorem states that different realizations of a Weyl system on Hilberts spaces of square-integrable functions on different Lagrangian subspaces of the same symplectic vector space are unitarily related, a conspicuous and well known example being the realization, in the case of
$T^{\ast}\mathbb{R}^{n}$ with coordinates $\left(  q^{i},p_{i}\right)  $ and
with the standard symplectic form, \ of the associated Weyl system on
square-integrable functions of the $q$'s or, alternatively, of the $p$'s, that
are related by the Fourier transform. \ In this sense the theorem is a
\textit{uniqueness }(up to unitary equivalence) theorem. We would like to
stress here that it is such if the linear structure (and symplectic form) are
assumed to be given once and for all.

As an example we shall consider here the case of a charged particle in a
constant magnetic field \cite{Zam} (and in the symmetric gauge) as described in the
previous section, reinstating Planck's constant in the appropriate places. We
can choose as Hilbert space that of the square-integrable functions on the
Lagrangian subspace defined by: $U^{i}=0,i=1,2$ (i.e. the subspace:
$u^{i}=-A^{i}\left(  q\right)  $ in the original coordinates).
Square-integrable wave functions will be denoted as $\psi\left(  Q^{1}%
,Q^{2}\right)  $ or $\psi\left(  Q\right)  $ for short.\ Then we can define
the Weyl operators:%
\begin{equation}
\fl \widehat{\mathcal{W}}(x,\pi)=\exp\left\{  \frac{i}{\hbar}\left[  x\widehat
{U}-\pi\widehat{Q}\right]  \right\}  =:\exp\left\{  \frac{i}{\hbar}\left[
x_{1}\widehat{U}^{1}+x_{2}\widehat{U}^{2}-\pi_{1}\widehat{Q}^{1}-\pi
_{2}\widehat{Q}^{2}\right]  \right\}  \label{weyl1}%
\end{equation}
acting on wavefunctions as:%
\begin{equation}
\left(  \widehat{\mathcal{W}}(x,\pi)\psi\right)  \left(  Q\right)
=\exp\left\{  -\frac{i}{\hbar}\pi\left(  Q+\frac{x}{2}\right)  \right\}
\psi\left(  Q+x\right)  \label{weyl2}.%
\end{equation}
Then: $\widehat{U}=-i\hbar\mathbf{\nabla}_{Q}$ while $\widehat{Q}$ acts as
the usual multiplication operator, i.e.: $(\widehat{Q}^{i}\psi)\left(
Q\right)  =Q^{i}\psi\left(  Q\right)  $. Eq. (\ref{weyl1}) can be rewritten
in a compact way as:%
\begin{equation}
\widehat{\mathcal{W}}(x,\pi)=\exp\left\{  \frac{i}{\hbar}\xi^{T}%
\mathbf{g}\widehat{X}\right\},
\end{equation}
where:%
\begin{equation}
\xi=\left\vert
\begin{array}
[c]{c}%
x\\
\pi
\end{array}
\right\vert ,\;\widehat{X}=\left\vert
\begin{array}
[c]{c}%
\widehat{U}\\
\widehat{Q}%
\end{array}
\right\vert
\end{equation}
and:%
\begin{equation}
\mathbf{g}=\left\vert
\begin{array}
[c]{cc}%
\mathbb{I}_{2\times2} & \mathbf{0}\\
\mathbf{0} & -\mathbb{I}_{2\times2}%
\end{array}
\right\vert.
\end{equation}
The dynamical evolution defines then the one-parameter family of Weyl
operators:%
\begin{equation}
\fl \widehat{\mathcal{W}}_{t}\left(  x,\pi\right)  =\widehat{\mathcal{W}}\left(
x\left(  t\right)  ,\pi\left(  t\right)  \right)  =\exp\left\{  \frac{i}%
{\hbar}\left[  x\left(  t\right)  \widehat{U}-\pi\left(  t\right)  \widehat
{Q}\right]  \right\}  \equiv\exp\left\{  \frac{i}{\hbar}\xi^{T}\left(
t\right)  \mathbf{g}\widehat{X}\right\}  \label{weyl3},%
\end{equation}
where:%
\begin{equation}
\xi\left(  t\right)  =\mathbb{F}\left(  t\right)  \xi.
\end{equation}
According to the standard procedure, this can be rewritten as:
\begin{equation}
\widehat{\mathcal{W}}_{t}\left(  x,\pi\right)  =\exp\left\{  \frac{i}{\hbar
}\left[  x\widehat{U}\left(  t\right)  -\pi\widehat{Q}\left(  t\right)
\right]  \right\}  =\exp\left\{  \frac{i}{\hbar}\xi^{T}\mathbf{g}\widehat
{X}\left(  t\right)  \right\},
\end{equation}
where:%
\begin{equation}
\eqalign{
\widehat{X}\left(  t\right)  =\widetilde{\mathbb{F}}\left(  t\right)
\widehat{X} \\
\widetilde{\mathbb{F}}\left(  t\right)  =\mathbf{g}\mathbb{F}\left(  t\right)
^{T}\mathbf{g} }
\end{equation}
and $\ \mathbb{F}\left(  t\right)  ^{T}$ denotes the transpose of the matrix
\ $\mathbb{F}\left(  t\right)  $. Explicitly:
\begin{equation}
\eqalign{ \fl
\widehat{U}^{1}\left(  t\right)  =\frac{1}{2}\widehat{U}^{1}(1+\cos\left(
Bt\right)  )-\frac{1}{2}\widehat{U}^{2}\sin\left(  Bt\right)  +\frac{B}
{4}\widehat{Q}^{1}\sin\left(  Bt\right)  -\frac{B}{4}\widehat{Q}^{2}\left(
1-\cos\left(  Bt\right)  \right) \\
\fl \widehat{U}^{2}\left(  t\right)  =\frac{1}{2}\widehat{U}^{1}\sin\left(
Bt\right)  +\frac{1}{2}\widehat{U}^{2}\left(  1+\cos\left(  Bt\right)
\right)  -\frac{B}{4}\widehat{Q}^{1}\left(  \cos\left(  Bt\right)  -1\right)
+\frac{B}{4}\widehat{Q}^{2}\sin\left(  Bt\right), }
\end{equation}
and:%
\begin{equation}
\eqalign{ \fl\widehat{Q}^{1}\left(  t\right)  =\frac{1}{B}\widehat{U}^{1}\sin\left(
Bt\right)  +\frac{1}{B}\widehat{U}^{2}\left(  \cos(Bt\right)  -1)-\frac{1}%
{2}\widehat{Q}^{1}(1+\cos\left(  Bt\right)  )+\frac{1}{2}\widehat{Q}^{2}%
\sin\left(  Bt\right) \\
\fl \widehat{Q}^{2}\left(  t\right)  =\frac{1}{B}\widehat{U}^{1}\left(
1-\cos\left(  Bt\right)  \right)  +\frac{1}{B}\widehat{U}^{2}\sin\left(
Bt\right)  -\frac{1}{2}\widehat{Q}^{1}\sin\left(  Bt\right)  -\frac{1}%
{2}\widehat{Q}^{2}(1+\cos\left(  Bt\right)  ) .}
\end{equation}

Now:%
\begin{equation}
\widehat{\mathcal{W}}_{t}\left(  x,\pi\right)  =\widehat{\mathcal{U}}\left(
t\right)  ^{\dag}\widehat{\mathcal{W}}\left(  x,\pi\right)  \widehat
{\mathcal{U}}\left(  t\right)  ;\;\widehat{\mathcal{U}}\left(
t\right)  =\exp\left\{  -\frac{it}{\hbar}\widehat{\mathcal{H}}\right\}
\end{equation}
and hence:
\begin{equation}
\widehat{Q}^{i}\left(  t\right)  =\widehat{\mathcal{U}}\left(  t\right)
^{\dag}\widehat{Q}^{i}\widehat{\mathcal{U}}\left(  t\right)
\end{equation}
and similarly for the $\widehat{U}^{i}$'s. Expanding in $t$ we find the
commutation relations:%
\begin{equation}
\eqalign{ 
\frac{i}{\hbar}\left[  \widehat{U}^{1},\widehat{\mathcal{H}}\right]  =\frac
{B}{2}\left(  \widehat{U}^{2}-\frac{B}{2}\widehat{Q}^{1}\right) \\
\frac{i}{\hbar}\left[  \widehat{U}^{2},\widehat{\mathcal{H}}\right]
=-\frac{B}{2}\left(  \widehat{U}^{1}+\frac{B}{2}\widehat{Q}^{2}\right)}
\end{equation}
and:%
\begin{equation}
\eqalign{
\frac{i}{\hbar}\left[  \widehat{Q}^{1},\widehat{\mathcal{H}}\right]  =-\left(
\widehat{U}^{1}+\frac{B}{2}\widehat{Q}^{2}\right) \\
\frac{i}{\hbar}\left[  \widehat{Q}^{2},\widehat{\mathcal{H}}\right]  =-\left(
\widehat{U}^{2}-\frac{B}{2}\widehat{Q}^{1}\right)}
\end{equation}
that, using the commutation relations: $\left[  \widehat{Q}^{i},\widehat{U}^{j}\right]  =i\hbar
\delta^{ij}$ are consistent with the Hamiltonian:%
\begin{equation}
\widehat{\mathcal{H}}=\frac{1}{2}\left\{  \left(  \widehat{U}^{1}+\frac{B}%
{2}\widehat{Q}^{2}\right)  ^{2}+\left(  \widehat{U}^{2}-\frac{B}{2}\widehat
{Q}^{1}\right)  ^{2}\right\},
\end{equation}
which is the quantum version of the Hamiltonian (\ref{Ham}).\\

In the general case, if two non-linearly related linear structures (and associated symplectic forms) are
available, then one can set up two different Weyl systems realized on two different Hilbert spaces.  Functions that are square-integrable in one setting need not be such in the other and
viceversa, and that because, as already remarked, the Jacobian of the
coordinate transformation is \textit{not} a constant. Morover, a  necessary
ingredient in the Weyl quantization program is the use of the (standard or
symplectic) Fourier transform. For the same reasons as outlined above, it is
also clear that, as already discussed, the two different linear structures
will define genuinely \textit{different}   Fourier transforms. 

In this way one can "evade" the uniqueness part of von Neumann's theorem. What the present
discussion is actually meant at showing is that there are assumptions, namely
that the linear structure (and symplectic form) are given once and for all and
are unique, that are implicitly assumed but not explicitly stated in the usual
formulations of the theorem, and that, whenever more structures are available,
the situation can be much richer and lead to genuinely and non-equivalent (in
the unitary sense) formulations of Quantum Mechanics. 

Let us illustrate these considerations by going back to the example of the $1D$ harmonic oscillator that has been discussed in section $1$. To quantize this system according to the Weyl scheme we have first of all to select a Lagrangian subspace ${\cal L}$  of $\mathbb{R}^2$ and a Lebesgue measure $d\mu$ on it defining then $L^2({\cal L},d\mu)$. When we endow $\mathbb{R}^2$ of the standard linear structure (\ref{lin1}) we chose ${\cal L} = \{ (q,0) \} $ and $d\mu = dq$. Alternatively, when we use the linear structure (\ref{lin2}), we take ${\cal L}^{\prime} = \{ (Q,0) \} $ and $d\mu = dQ$. Notice that ${\cal L}$ and ${\cal L}^{\prime}$ are the same  subset of $\mathbb{R}^2$, defined by the conditions $ P=p=0 $  and with the coordinates related by the relation $Q= q K(r=|q|) $. Nevertheless the two Hilbert spaces $L^2({\cal L},d\mu)$  and $L^2({\cal L}^{\prime},d\mu ^{\prime})$  are not related via a unitary map since the Jacobian of the coordinate transformations is not constant: $ d\mu = (1+3\lambda Q^2) 
d\mu^{\prime}$. 

As a second step in the Weyl scheme, we construct in  $L^2({\cal L},d\mu)$ the operator   $\hat{U}(\alpha)$:
\begin{equation}
\left( \hat{U}(\alpha) \psi \right) (q)  = e^{i\alpha q/\hbar } \psi (q) \; , \; \psi (q) \in L^2({\cal L},d\mu),
\end{equation}
whose generator is  $\hat{x} = q$, and the operator  $\hat{V}(h)$:
\begin{equation}
\left( \hat{V}(h) \psi \right) (q)  = \psi (q+h) \; \psi (q) \in L^2({\cal L},d\mu),
\end{equation}
which is generated by  $ \hat{\pi} = - i \hbar \partial/\partial q$, and implements the translations defined by the linear structure (\ref{lin1}). The quantum Hamiltonian can be written as $H= \hbar \left(a^\dagger a + \frac{1}{2}\right)$ where $a = (\hat{x}+i \hat{\pi})/ \sqrt{2}\hbar$ (here the adjoint is taken with respect to the complex structure defined by the Lebesgue measure $dq$). \\
Similar expressions hold in $L^2({\cal L}^{\prime},d\mu ^{\prime})$ for $\hat{x}^{\prime}$, $ \hat{\pi}  ^{\prime} $  and $\hat{U}^{\prime} (\alpha)$, $\hat{V}^{\prime}(h) $. Notice that, as seen as operators in the former Hilbert space, $\hat{V}^{\prime}(h) $ implements translations with respect to the linear structure (\ref{lin2}):
\begin{equation}
(\hat{V}^{\prime}(h) \psi)(q) = \psi (q +_{(K)} h).
\end{equation}
Now the quantum Hamiltonian is $H^{\prime}= \hbar \left(A{^\dagger\prime} A + \frac{1}{2}\right)$ with $A = (\hat{x}^{\prime}+i \hat{\pi}^{\prime})/ \sqrt{2}\hbar$, where now the adjoint is taken with respect to the complex structure defined by the Lebesgue measure $dQ$\footnote{A direct calculation shows that $a^{\dagger} = \frac{1}{\sqrt{2}\hbar} \left( q -\hbar \frac{\partial}{\partial q}\right)$ whereas 
$a^{\dagger\prime} =  \frac{1}{\sqrt{2}\hbar} \left[q -\hbar \frac{\partial}{\partial q} -  \frac{ 6\hbar \lambda K(r) q}{(1+3\lambda K(r)^2 q^2)^2} \right]$. Also $A{^\dagger\prime} =  \frac{1}{\sqrt{2}\hbar} \left[ K(r)q -\hbar (1+3\lambda K(r)^2q^2) \frac{\partial}{\partial q}\right]$. We notice that the transformation relating $a^{\dagger}$ and  $A{^\dagger\prime} $  is not of the type considered in in the second reference of \cite{Vent}.}.

Finally we recall that, following the Weyl-Wigner-Moyal program \cite{Rev}, one defines an
``inverse" mapping of (actually Hilbert-Schmidt) operators onto
square-integrable functions in phase space endowed with a non-commutative
``$\ast$-product", the Moyal product \cite{Moy}. The
Moyal product is defined as:%
\begin{equation}\label{moyal}
\left(  f\ast g\right)  \left(  q,p\right)  =f\left(  q,p\right)  \exp\left\{
\frac{i\hbar}{2}\left[  \overleftarrow{\frac{\partial}{\partial q}%
}\overrightarrow{\frac{\partial}{\partial p}}-\overleftarrow{\frac{\partial
}{\partial p}}\overrightarrow{\frac{\partial}{\partial q}}\right]  \right\}
g\left(  q,p\right).
\end{equation}
It defines in turn the Moyal bracket:%
\begin{equation}
\left\{  f,g\right\}  _{M}=:\frac{1}{i\hbar}\left(  f\ast g-g\ast f\right)
\end{equation}
and:
\begin{equation}
\left\{  f,g\right\}  _{M}=\left\{  f,g\right\}  _{\omega}+\mathcal{O}\left(
\hbar^{2}\right),
\end{equation}
where $\left\{  .,\right\}_{\omega}$ is the Poisson bracket defined by the
symplectic form $\omega$, and similarly with the use of the
second  (i.e., $\left(  \partial/\partial Q,\partial/\partial P\right)  $) linear structure. Different (and not unitarily equivalent) Weyl systems will lead to different Moyal products and brackets, and to different Poisson brackets in
the classical limit. 

For example, in the case of the  $1D$ harmonic oscillator one has Eqn. (\ref{moyal}) for the ordinary Moyal product and,
\begin{equation}
\left(  f\ast_K g\right)  \left(  Q,P\right)  =f\left(  Q,P\right)  \exp\left\{
\frac{i\hbar}{2}\left[  \overleftarrow{\frac{\partial}{\partial Q}%
}\overrightarrow{\frac{\partial}{\partial P}}-\overleftarrow{\frac{\partial
}{\partial P}}\overrightarrow{\frac{\partial}{\partial Q}}\right]  \right\}
g\left(  Q,P\right),
\end{equation}
which define the corresponding Moyal brackets $\left\{  f,g\right\}  _{M}$ and $\left\{  f,g\right\} _ {M_K}$. It is not difficult to check that, since
\begin{equation}
\left[  \overleftarrow{\frac{\partial}{\partial Q}%
}\overrightarrow{\frac{\partial}{\partial P}}-\overleftarrow{\frac{\partial
}{\partial P}}\overrightarrow{\frac{\partial}{\partial Q}}\right]   = D \, 
\left[  \overleftarrow{\frac{\partial}{\partial Q}%
}\overrightarrow{\frac{\partial}{\partial P}}-\overleftarrow{\frac{\partial
}{\partial P}}\overrightarrow{\frac{\partial}{\partial Q}}\right]  ,
\end{equation} 
 eq.  (\ref{fg}) will hold in the classical limit $\hbar \to 0$.

\appendix

\section{Further examples of "exported" linear structures}

\begin{itemize}
\item  Relativistic addition of velocities.   Let $E=\mathbb{R}$, $M=\left(  -1,1\right)  $ and:%
\begin{equation}
\phi:x\rightarrow X=:\tanh x.
\end{equation}
Then:%
\begin{equation}
\lambda \cdot_{\left(  \phi\right)  } X=\tanh\left(  \lambda\tanh
^{-1}\left(  X\right)  \right)
\end{equation}
and:%
\begin{equation}
\eqalign{ \lambda \cdot_{\left(  \phi\right)  } \left(  \lambda^{\prime } \cdot_{\left(  \phi\right)  } X\right)  &=\lambda \cdot_{\left(  \phi\right)  } \tanh\left(  \lambda^{\prime}\tanh^{-1}\left(  X\right)
\right)  =\\
 &=\tanh\left(  \lambda\lambda^{\prime}\tanh^{-1}\left(  X\right)  \right)
=\left(  \lambda\lambda^{\prime}\right)  \cdot_{\left(  \phi\right)  } X, }
\end{equation}
while:%
\begin{equation}
X +_{\left(  \phi\right)  } Y=\tanh\left(  \tanh^{-1}\left(  X\right)
+\tanh^{-1}\left(  Y\right)  \right)  =\frac{X+Y}{1+XY},
\end{equation}
which is nothing but the one-dimensional relativistic law (in appropriate
units) for the addition of velocities. It is also simple to prove that:%
\begin{equation}%
\eqalign{
\left(  X +_{\left(  \phi\right)  } Y\right)  +_{\left(  \phi\right)  }  Z &=\tanh\left(  \tanh^{-1}\left(  X +_{\left(  \phi\right)  } Y\right)  +\tanh^{-1}\left(  Z\right)  \right)  =\\
&=\tanh\left(  \tanh^{-1}X+\tanh^{-1}\left(  Y\right)  +\tanh^{-1}\left(
Z\right)  \right)}
\end{equation}
i.e. that:%
\begin{equation}
\left(  X +_{\left(  \phi\right)  } Y\right)  +_{\left(  \phi\right)  } Z=X +_{\left(  \phi\right)  } \left(  Y +_{\left(  \phi\right)  } Z\right).
\end{equation}

Explicitly:%
\begin{equation}
X +_{\left(  \phi\right)  } Y +_{\left(  \phi\right)  }  Z=\frac{X+Y+Z+XYZ}{1+XY+XZ+YZ}.%
\end{equation}

The\ mapping (\ref{Liouville0}) is now:%
\begin{equation}
X\left(  t\right)  =\tanh\left(  e^{t}\tanh^{-1}\left(  X\right)  \right)
\end{equation}
and we obtain, for the Liouville field on $\left(  -1,1\right)  $:%
\begin{equation}
\Delta\left(  X\right)  =\left(  1-X^{2}\right)  \tanh^{-1}\left(  X\right)
\frac{\partial}{\partial X}%
\end{equation}
and $\Delta\left(  X\right)  =0$ for $X=0$.

\item Another similar example involves $E=\mathbb{R},M=\mathbb{R}^{+}=\left(
0,+\infty\right)  $ and:%
\begin{equation}
\phi=x\rightarrow X=\exp\left(  x\right).
\end{equation}
Then one can see easily that:%
\begin{equation}
X \cdot_{\left(  \phi\right)  } X^{\prime}=XX^{\prime}%
\end{equation}
and:%
\begin{equation}
\lambda \cdot_{\left(  \phi\right)  } X=X^{\lambda}.
\end{equation}

In this way:
\begin{equation}
X\left(  t\right)  =\phi\left(  e^{t}\phi^{-1}(X)\right)  =X^{e^{t}}%
=\exp\left[  e^{t}\ln\left(  X\right)  \right]
\end{equation}
and one finds the "adapted" Liouville field:%
\begin{equation}
\Delta\left(  X\right)  =X\ln\left(  X\right)  \frac{\partial}{\partial X}.%
\end{equation}
Notice that here the fixed point of the Liouville field is $X=1=\phi\left(
0\right)  $.

\item \ (In this example $\phi$ is a homeomorphism and not a diffeomorphism).
Let $E=M=\mathbb{R}$ and:%
\begin{equation}
\phi:x\rightarrow X=x^{3}.%
\end{equation}
Then:%
\begin{equation}
X +_{(\phi)} Y=\left(  \sqrt[3]{X}+\sqrt[3]{Y}\right)  ^{3}%
\end{equation}
and:%
\begin{equation}
\lambda \cdot_{(\phi)}X=\lambda^{3}X.
\end{equation}
The proof that \ (\ref{property3}) and (\ref{property4}) are satisfied is
elementary and will be omitted.

\item As a last example, we can consider the inverse stereographic projection
of $\mathbb{R}^{2}$ onto the Riemann sphere:%
\begin{equation}
\Phi:\mathbb{R}^{2}\rightarrow\mathbb{S}^{2}-\left\{  0,0,1\right\}
\end{equation}
by:%
\begin{equation}
\fl \Phi:z\rightarrow X=\left\{  x_{1},x_{2},x_{3}\right\}  ,x_{1}=\frac
{z+\overline{z}}{\left\vert z\right\vert ^{2}+1},x_{2}=\frac{z-\overline{z}%
}{i(\left\vert z\right\vert ^{2}+1)},x_{3}=\frac{\left\vert z\right\vert
^{2}-1}{\left\vert z\right\vert ^{2}+1} \label{Riemann1},%
\end{equation}
that inverts to:%
\begin{equation}
z=\Phi^{-1}\left(  X\right)  =\frac{x_{1}+ix_{2}}{1-x_{3}} \label{Riemann2}.%
\end{equation}
Multiplication by a constant results in:%
\begin{equation}
\lambda \cdot_{(\Phi)}X=X^{\prime}%
\end{equation}
with $X^{\prime}=\left\{  x_{1}^{\prime},x_{2}^{\prime},x_{3}^{\prime
}\right\}  $ and:%
\begin{equation}
\fl x_{1}^{\prime}=\frac{2\lambda x_{1}}{\lambda^{2}+1+x_{3}\left(  \lambda
^{2}-1\right)  },x_{2}^{\prime}=\frac{2\lambda x_{1}}{\lambda^{2}%
+1+x_{3}\left(  \lambda^{2}-1\right)  },x_{3}^{\prime}=\frac{\lambda
^{2}-1+x_{3}\left(  \lambda^{2}+1\right)  }{\lambda^{2}+1+x_{3}\left(
\lambda^{2}-1\right)  },%
\end{equation}
$\left(  x_{1}^{\prime}\right)  ^{2}+\left(  x_{2}^{\prime}\right)
^{2}+\left(  x_{3}^{\prime}\right)  ^{2}=1$. Again: $X^{\prime}=X$ \ for
$\lambda=1$, while, for $\lambda=0,X^{\prime}=\left\{  0,0,-1\right\}  $.
Moreover, for $\lambda\rightarrow\infty$, $x_{1,2}^{\prime}\rightarrow0$
while: $x_{3}^{\prime}\rightarrow1$. \ Setting then: $\lambda=e^{t}$ and
taking derivatives, we obtain:%
\begin{equation}
\frac{dx_{1}}{dt}|_{t=0}=-x_{1}x_{3},\frac{dx_{2}}{dt}|_{t=0}=-x_{2}x_{3}
\label{equation1}%
\end{equation}
and:%
\begin{equation}
\frac{dx_{3}}{dt}=1-x_{3}^{2} .\label{equation2}%
\end{equation}
This defines the vector field:%
\begin{equation}
\Delta=-x_{1}x_{3}\frac{\partial}{\partial x_{1}}-x_{2}x_{3}\frac{\partial
}{\partial x_{2}}+\left(  1-x_{3}^{2}\right)  \frac{\partial}{\partial x^{3}}
\label{Liouvillesphere}%
\end{equation}
and it is easy to check that:
\begin{equation}
\mathcal{L}_{\Delta}\left(  \left(  x_{1}\right)  ^{2}+\left(  x_{2}\right)
^{2}+\left(  x_{3}\right)  ^{2}\right)  =0,
\end{equation}
i.e. that $\Delta$ is indeed tangent to the sphere.

Switching to spherical polar coordinates:%
\begin{equation}
x_{1}=\sin\theta\cos\phi,x_{2}=\sin\theta\sin\phi,x_{3}=\cos\theta,
\end{equation}
the equations of motion (\ref{equation1},\ref{equation2})
take the simple form:%
\begin{equation}
\frac{d\theta}{dt}=-\sin\theta,\frac{d\phi}{dt}=0 \label{equation3}%
\end{equation}
and hence the field (\ref{Liouvillesphere}) become simply:%
\begin{equation}
\Delta=-\sin\theta\frac{\partial}{\partial\theta},%
\end{equation}
which has a fixed point at $\theta=\pi$.
\end{itemize}

Explicitly, eq.ns (\ref{equation3}) integrate to:%
\begin{equation}
\tan\left(  \frac{\theta\left(  t\right)  }{2}\right)  =\tan\left(
\frac{\theta}{2}\right)  e^{-t},\phi\left(  t\right)  =const.=\phi
\end{equation}
$\left(  \theta\left(  0\right)  =\theta,\phi\left(  0\right)  =\phi\right)
$, and $\theta$ flows towards the "North Pole" $\theta=0$ when $t\rightarrow
+\infty$ and towards the "South Pole" $\theta=\pi$ when $t\rightarrow-\infty$.

Polar coordinates make all the calculations easier. Representing $z$ as:
$z=\rho\exp\left(  i\varphi\right)  $ and $X$ with the polar angles $\theta$
and $\phi$, the maps (\ref{Riemann1}) and (\ref{Riemann2}) become simply:%
\begin{equation}%
\eqalign{
\phi=\varphi\\
\sin\theta=\frac{2\rho}{\rho^{2}+1},\cos\theta=\frac{\rho^{2}-1}{\rho^{2}+1} }
\label{map7}%
\end{equation}
and:%
\begin{equation}%
\eqalign{
\varphi=\phi\\
\rho=\cot\left(  \theta/2\right) }
\end{equation}
respectively. \\
Then, given $X=\left(  \theta,\phi\right)  ,X^{\prime}=\left(
\theta^{\prime},\phi^{\prime}\right)$:
\begin{equation}
\Phi^{-1}\left(  X\right)  +\Phi^{-1}\left(  X^{\prime}\right)  =\rho
\exp\left(  i\varphi\right),
\end{equation}
with:%
\begin{equation}
\rho=\sqrt{\cot^{2}\left(  \theta/2\right)  +\cot^{2}\left(  \theta^{\prime
}/2\right)  +2\cot\left(  \theta/2\right)  \cot\left(  \theta^{\prime
}/2\right)  \cos\left(  \phi-\phi^{\prime}\right)  } \label{rho}%
\end{equation}
and:%
\begin{equation}
\fl \cos\varphi=\frac{\cot\left(  \theta/2\right)  \cos\phi+\cot\left(
\theta^{\prime}/2\right)  \cos\phi^{\prime}}{\rho},\sin\varphi=\frac
{\cot\left(  \theta/2\right)  \sin\phi+\cot\left(  \theta^{\prime}/2\right)
\sin\phi^{\prime}}{\rho} \label{phi2}.%
\end{equation}
It follows that:%
\begin{equation}
X +_{(\Phi)} X^{\prime}=\left(  \theta,\phi \right)  +_{(\Phi)} \left(  \theta^{\prime},\phi^{\prime}\right)  =\left(  \theta
^{\prime\prime},\phi^{\prime\prime}\right),
\end{equation}
where $\phi^{\prime\prime}=\varphi$, with $\varphi$ given by Eqn.(\ref{phi2})
and:%
\begin{equation}
\sin\theta^{\prime\prime}=\frac{2\rho}{\rho^{2}+1},\cos\theta^{\prime\prime
}=\frac{\rho^{2}-1}{\rho^{2}+1},%
\end{equation}
with $\rho$ given now by Eqn.(\ref{rho}). In particular, if $\phi=\phi
^{\prime}$, then $\phi^{\prime\prime}=\phi$ and:%
\begin{equation}%
\eqalign{
\sin\theta^{\prime\prime}=2\frac{\cot\left(  \theta/2\right)  +\cot\left(
\theta^{\prime}/2\right)  }{\left(  \cot\left(  \theta/2\right)  +\cot\left(
\theta^{\prime}/2\right)  \right)  ^{2}+1},\\
\cos\theta^{\prime\prime}=\frac{\left(  \cot\left(  \theta/2\right)
+\cot\left(  \theta^{\prime}/2\right)  \right)  ^{2}-1}{\left(  \cot\left(
\theta/2\right)  +\cot\left(  \theta^{\prime}/2\right)  \right)  ^{2}+1} .}
\end{equation}

Concerning multiplication by a (real) constant, we have:%
\begin{equation}
\lambda \cdot_{(\Phi)} X=\lambda \cdot_{(\Phi)} \left(
\theta,\phi\right)  =\left(  \theta^{\prime},\phi\right),
\end{equation}
with:%
\begin{equation}
\sin\theta^{\prime}=\frac{2\lambda\cot\left(  \theta/2\right)  }{\lambda
^{2}\cot^{2}\left(  \theta/2\right)  +1},\cos\theta^{\prime}=\frac{\lambda
^{2}\cot^{2}\left(  \theta/2\right)  -1}{\lambda^{2}\cot^{2}\left(
\theta/2\right)  +1}.
\end{equation}
Here too, for $\lambda\rightarrow0,+\infty$,  $\theta^{\prime}$ flows
towards the South Pole and the North Pole respectively.

\end{document}